\documentstyle[11pt,a4wide]{article}
\def\be{\begin{equation}}
\def\ee{\end{equation}}
\begin{document}
\centerline{\Large  \bf Schwarzschild and Synge once again}
%\vspace*{1cm}
\bigskip
\centerline{\large Hans - J\"urgen  Schmidt}
\medskip
\centerline{http://www.physik.fu-berlin.de/\~{}hjschmi, \ 
e-mail: hjschmi@rz.uni-potsdam.de}
\medskip
\centerline{Institut f\"ur Mathematik, Universit\"at
Potsdam, Am Neuen Palais 10,  D-14469 Potsdam, Germany}
\bigskip
\begin{abstract}
We complete the historical overview about the geometry of a 
Schwarzschild black hole at its horizon by emphasizing the 
contribution  made by  J. L. Synge in 1950 to its clarification. 
\end{abstract}

KEY: Schwarzschild solution; Synge; black hole; horizon

\section{Introduction}

P. Florides raised the point\footnote{in his plenary lecture
at the Conference GR17 in Dublin, July 18-24, 2004} 
that the contribution of Synge for the clarification 
of the geometry of the Schwarzschild horizon is still 
not adequately known to the scientific community. 
It is the purpose of the present note  to fill this gap. 

\bigskip

In 1916, the paper [1] appeared whose resulting solution is
 now known as the Schwarzschild solution. The most 
 well-known early discussion about its horizon is 
 [2] and [3]; there, Einstein erroneously believed to have proven that no 
particle can cross the horizon of a Schwarzschild black hole.
 In [4], however, by use of the coordinates later named
Eddington-Finkelstein coordinates, examples of particles
 crossing he horizon could be constructed. But 
the complete geometry near the horizon was still unclear that time.

\section{Heckmann's view of Synge's paper}

In 1951, O. Heckmann from Hamburg/Germany published
the following notes [5]\footnote{Translation from the German
original by H.-J. Schmidt} about the Synge-paper [6] from 1950: 
``The known metric
$$
ds^2 = e^{- \lambda} dr^2 + r^2 ( d \theta^2 + \sin^2 \theta d \varphi^2 )
 - e^{\lambda} dt^2
$$
with $\lambda = \ln (1-a/r)$; $a = {\rm const.} > 0$, is assumed to be
a singular one at $a=r$
in almost every paper about General Relativity Theory. 
The paper by Synge shows however, that this singularity can be removed 
by a suitable choice of coordinates. The time-like geodesics (paths of 
particles) and the null-like geodesics (light rays) are discussed in details,
and the result is that a particle can indeed cross the place $r=a$; then, 
after a finite eigentime, it reaches $r=0$ at light velocity. The paper by Synge 
contains many interesting results all of them being discussed in details."
 \footnote{Translator's note: Nowadays the parameter $a$
is almost always replaced by $2m$.}

\section{Discussion}

In [7], Ehlers wrote: ``An outstanding achievement of Synge's was the
first complete analytic extension of the Schwarzschild field.", but he did not
give the related source. In fact, it is Synge [6] entitled ``The gravitational 
field of a particle", and this seems really  to be the first among 
many sources of this result.

For more details see [8]; there it is also proposed to replace the
notion ``inside the horizon" by ``after the horizon". This is not 
only done from the formal point of view because $r$ is time-like at 
$r < 2m$, but also from  the mental point of view: 
A result, that a particle cannot leave the region ``after the
horizon", is not a strange new behaviour from relativity theory, 
but nothing but a variant of the well-known classical result 
that it is not possible to return from the future back to the past. 

{\bf  Note added:} Further details can be found in [9]. In
 [10], Lemaitre (1933) showed, though not  the complete analytic 
extension of the Schwarzschild solution, (similarly as
 Eddington [4] did in 1924), that the surface $r=2m$ is
a  fictituous singularity. 
   
\bigskip

{\Large \bf Acknowledgement}

I am grateful to A. Gsponer and S. Robertson for valuable comments to
the first version of this paper. 

\bigskip

{\Large \bf   References}

[1] K. Schwarzschild: Sitzungsber. Preuss. Akad. Wiss., Phys.-Math. 
Klasse (1916) 189. Translation from the German original in: 
Gen. Rel. Grav. {\bf 35} (2003)  951 by S. Antoci und A. Loinger. 
 Editor's Note to this paper: S. Antoci und D.-E. Liebscher: 
Gen. Rel. Grav. {\bf 35} (2003)  945. 

[2] E. Trefftz, Mathem. Ann. {\bf 86} (1922) 317.

[3] A. Einstein: Sitzungsber. Preuss. Akad. Wiss., Phys.-Math. 
Klasse  (1922)  448.

[4] A. S. Eddington: Nature {\bf 113} (1924) 192. 

[5] O. Heckmann: Zentralbl. Math. {\bf 36} (1951) 426.

[6] J. L. Synge: Proc. Irish Acad. A {\bf 53} (1950) 83-114.

[7]  J. Ehlers:  Gen. Rel. Grav. {\bf 34} (2002)  2171. 

[8] H.-J. Schmidt: Im Zwischenreich der Bilder, Eds.:  R. Jacobi, 
B. Marx, G. Strohmaier-Wiederanders, EVA  Leipzig (2004), page 267
(in German language).

[9] A. Gsponer, preprint physics/0408100.

[10] G. Lemaitre, Ann. Soc. Sc. Brux. A {\bf 53} (1933) 51, reprinted
 in Gen. Rel. Grav. {\bf 29} (1997)  641. 

\end{document}